\documentclass[reprint,amsmath,amssymb,aps,superscriptaddress,prl,nobibnotes]{revtex4-2}

\usepackage{graphicx}
\usepackage{dcolumn}
\usepackage{bm}

\usepackage[utf8]{inputenc}
\usepackage[T1]{fontenc}
\usepackage{mathptmx}
\usepackage{etoolbox}
\usepackage{xcolor}

\usepackage[colorlinks=true,allcolors=blue]{hyperref}

\begin{document}

\title[]{A scanning cavity for large area coherent coupling in atom interferometry}
\author{D. O. Sabulsky}
\affiliation{\href{https://ror.org/031ah7413}{Artemis}, CNRS, Observatoire de la C\^ote d'Azur, Universit\'e C\^ote d'Azur, 06304 Nice, France}
\affiliation{\href{https://ror.org/03hgsyv65}{LTE}, Observatoire de Paris, Universit{\'e} PSL, Sorbonne Universit{\'e}, Universit{\'e} de Lille, LNE, CNRS, 61 avenue de l’Observatoire, 75014 Paris, France}
\affiliation{\href{https://ror.org/03vdg7e33}{LP2N}, Laboratoire Photonique, Num\'erique et Nanosciences, Universit\'e de Bordeaux-IOGS-CNRS, Talence, France.}
\author{Q. Beaufils}
\affiliation{\href{https://ror.org/03hgsyv65}{LTE}, Observatoire de Paris, Universit{\'e} PSL, Sorbonne Universit{\'e}, Universit{\'e} de Lille, LNE, CNRS, 61 avenue de l’Observatoire, 75014 Paris, France}
\author{A. Bertoldi}
\affiliation{\href{https://ror.org/03vdg7e33}{LP2N}, Laboratoire Photonique, Num\'erique et Nanosciences, Universit\'e de Bordeaux-IOGS-CNRS, Talence, France.}
\author{A. Landragin}
\email{arnaud.landragin@obspm.fr}
\affiliation{\href{https://ror.org/03hgsyv65}{LTE}, Observatoire de Paris, Universit{\'e} PSL, Sorbonne Universit{\'e}, Universit{\'e} de Lille, LNE, CNRS, 61 avenue de l’Observatoire, 75014 Paris, France}
\author{P. Bouyer}
\affiliation{\href{https://ror.org/05wag9297}{Van der Waals-Zeeman Institute}, Institute of Physics, University of Amsterdam, Science Park 904, 1098 XH, Amsterdam, The Netherlands.}
\affiliation{\href{https://ror.org/00zq3ce72 }{QuSoft}, Science Park 123, 1098 XG, Amsterdam, The Netherlands.}
\affiliation{\href{https://ror.org/02c2kyt77}{Eindhoven University of Technology}, P.O. Box 513, 5600 MB, Eindhoven, The Netherlands.}
\author{B. Canuel}
\email{benjamin.canuel@institutoptique.fr}
\affiliation{\href{https://ror.org/03vdg7e33}{LP2N}, Laboratoire Photonique, Num\'erique et Nanosciences, Universit\'e de Bordeaux-IOGS-CNRS, Talence, France.}

\date{\today}

\begin{abstract}

Optical cavities can enhance atom–light coupling in atom interferometry, but free-fall geometries require large transverse modes. Marginally stable resonators provide such modes, yet aberrations and alignment imperfections turn their near-degenerate response into detuning-dependent transverse ring modes, strongly limiting the usable interaction volume under static excitation. Here, we turn this limitation into a resource. By chirping the interrogation frequency across the cavity resonance while dynamically shaping the input amplitude, we map laser detuning onto the transverse position and scan the cavity-enhanced coupling across the atomic cloud. In a cavity-enhanced Bragg diffraction experiment with $^{87}$Rb atoms, this produces a nearly uniform effective interaction region of about 15 mm and increases the transfer efficiency from about 5\% to 28\%. We also demonstrate phase-coherent operation of a three-pulse atom interferometer using chirped cavity-enhanced Bragg pulses. These results establish scanning-cavity interrogation as a route to large-area, cavity-enhanced atom interferometry.
\end{abstract}

\maketitle

\section*{Introduction}
Atom interferometry relies on the interactions between atoms and light that require precise control of the phase and intensity of the lasers that manipulate the matter-waves. Furthermore, when aiming for high sensitivity, high-intensity laser beams with a large area are required to allow for a long interaction time\cite{Dickerson2013,schuldt2015,Gautier2022} as well as large momentum transfer pulses \cite{Chiow2011,gebbe2021,wilkason2022,Rodzinka2024}.
Optical cavities are attractive in this context and have already demonstrated their potential in different ways \cite{Hamilton2015,Greve2022,Panda2024,Sabulsky2024}
and are expected to be a key tool for large base-line atom interferometry antennas \cite{Canuel2018,Zhan2019,Canuel2020}. In particular, marginally stable resonators can support large transverse beams \cite{Mielec2020} and therefore have emerged as a promising route toward cavity-enhanced atom interferometry.

However, near-degeneracy comes at a significant cost: in practice, aberrations and alignment imperfections strongly distort the intra-cavity field \cite{Arnaud1969}, such that the cavity response is no longer described by a single broad, homogeneous mode. Instead, the resonance becomes asymmetric and the resonant field develops a detuning-dependent transverse ring structure \cite{Riou2017}. 
In cavity-enhanced atom optics, this spatial structure restricts the usable interaction region and can introduce phase gradients across the atomic cloud, thereby limiting both the diffraction efficiency and the interferometric contrast \cite{Sabulsky2024}.

Here, we show that this apparent limitation can instead be used as a resource. Given that the cavity resonance condition depends on transverse position, the laser detuning maps directly onto the radius of the resonant field. By scanning the interrogation frequency across the cavity resonance while shaping the input amplitude, we dynamically scan the atom-light coupling across the cloud and synthesize an effective extended interaction region. In this way, the strongly inhomogeneous instantaneous cavity field is converted into a nearly homogeneous effective coupling over the scanned area. We present our implementation of this method in a cavity-enhanced Bragg diffraction experiment with $^{87}$Rb atoms. Compared to the case of static excitation \cite{Sabulsky2024}, we demonstrate that chirped interrogation significantly increases both the diffraction efficiency and the atom interferometer contrast.

\section*{Experimental system and cavity response}

\begin{figure*}
\centering
\includegraphics[width=1\linewidth]{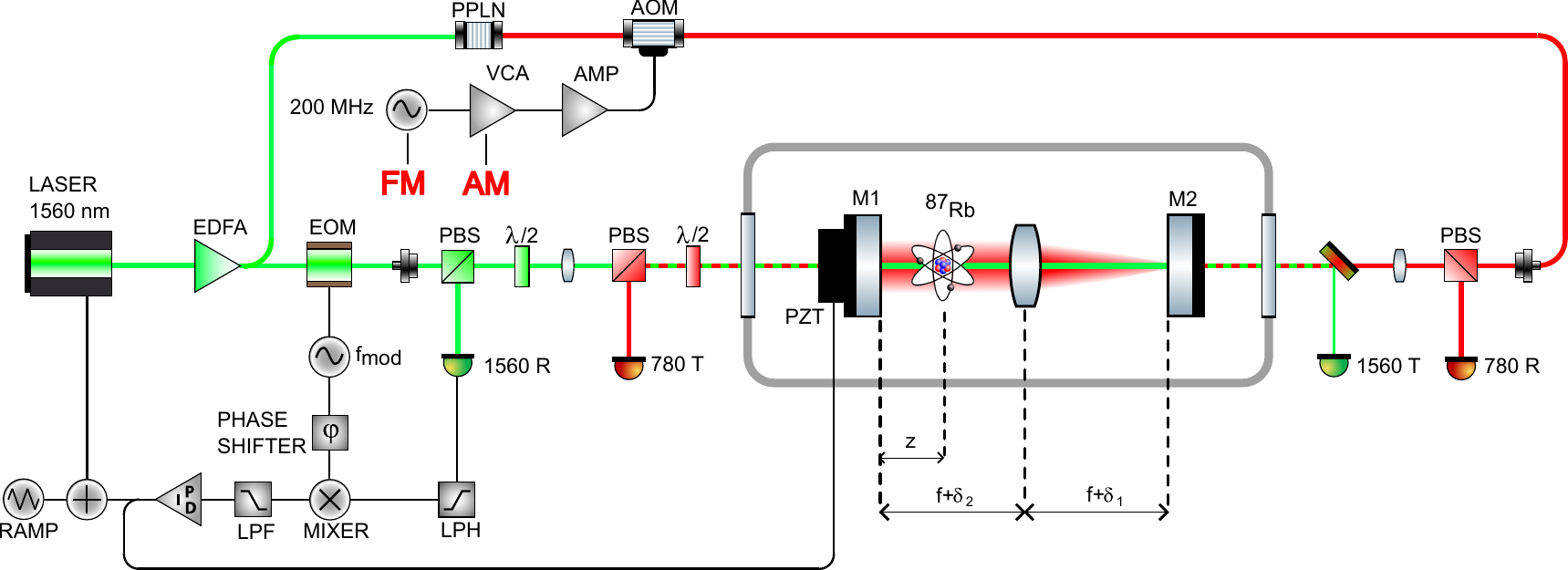}
\caption{\textbf{Schematic of the experimental setup.}
A marginally stable optical cavity ($L=2f$) is driven at 1560~nm (green path) and probed at 780~nm (red path). The 1560~nm laser is locked to the cavity using a PDH scheme, while the detuning (FM) and amplitude (AM) of the 780~nm light is controlled via  the RF signal driving an AOM. The atomic cloud is located at distance $z$ from mirror $M_1$.}
\label{fig:fig1}
\end{figure*}

A cold ensemble of $^{87}$Rb atoms at $2.5~\mu$K is launched vertically and interrogated near the apogee of its trajectory. After state preparation and velocity selection, a few $10^5$ atoms in the hyperfine ground state $F=1,m_{\mathrm{F}}=0$, with a momentum width of $0.2 \times v_{\mathrm{rec}}$, are addressed by a three-pulse horizontal Bragg interferometer. After interrogation, the transition probability is measured using a momentum-selective detection \cite{Beaufils2022}. During interrogation, the cloud has a transverse FWHM diameter of about $15.6$~mm, much larger than the resonant cavity field. Under these conditions, static excitation produces a low Bragg transfer efficiency because only a small central fraction of the ensemble is efficiently driven \cite{Sabulsky2024}.

The experiment is based on a marginally stable optical cavity operated in ultrahigh vacuum \cite{Sabulsky2024,Junca2022.PhD} (Fig.~\ref{fig:fig1}). The cavity consists of two flat mirrors separated by $L=2f$, with a biconvex lens of focal length $f=40$~cm inserted between them. Small longitudinal offsets from the focal planes determine the departure from perfect degeneracy. The measured finesse is $F=200$ corresponding to an optical gain of $G=38$, presently limited by aberrations of the intra-cavity lens . \textcolor{black}{Its determination is realized following the methods presented in \cite{Riou2017,Mielec2020}, and based on the measurement of the power transmitted by the cavity and of each mirror}. The resonant field has a FWHM of $\sim 4$~mm , significantly smaller than the expected Gaussian beam of $\sim 15$~mm FWHM  that should resonate for an ideal cavity of this geometry.

The cavity is length-frequency stabilized at 1560~nm and probed at 780 nm for single-frequency Bragg diffraction \textcolor{black}{as detailed in \cite{Sabulsky2024}}. The cavity-laser differential frequency is stabilized via Pound--Drever--Hall (PDH) \cite{Long2007} \textcolor{black}{on the blue-shifted sideband of the IR seed laser, green path in Fig.~\ref{fig:fig1}, generated by an electro-optic phase modulator (EOM)\textcolor{black}{. The locking system uses the blue-shifted sideband as the frequency reference to apply a slow feedback to a piezoelectric actuator (PZT) mounted on one of the cavity mirrors, while benefiting from the fast corrections provide by the laser current. The adjustment of the EOM's driving frequency allows the effective resonance frequency of the 780 nm  Bragg beam (red path in Fig.~\ref{fig:fig1})} to be controlled independently. This NIR wavlength is} obtained by amplification and subsequent frequency-doubling of the 1560 nm carrier in a PPLN crystal, is blue-detuned by approximately $3.4$~GHz from the D2 transition of $^{87}$Rb and is used to implement Bragg diffraction. The same beam is detuned from the cavity resonance by varying the driving frequency of an acousto-optic modulator (AOM). The beam amplitude is independently controlled through a voltage-controlled RF attenuator (VCA). This architecture provides independent access to the two ingredients of our protocol: a frequency chirp with amplitude shaping.

In this cavity, the resonance condition depends both on laser detuning $\delta\nu$ from the nominal cavity resonance and on the radial position $r$ with respect to the configuration axis:
\begin{equation}
I(\delta\nu,r)=
\frac{I_0}{1+4F^2\!\left(
\frac{\delta\nu}{\Delta\nu_{\mathrm{FSR}}}
-
\frac{\delta_1 r^2}{\lambda f^2}
\right)^2}
\,\exp\!\left(-\frac{2r^2}{\omega_0^2}\right),
\label{eq:static}
\end{equation}

\noindent where $\Delta\nu_{\mathrm{FSR}}$ is the free spectral range of the cavity and $\omega_0$ is the waist of the injected Gaussian beam; \textcolor{black}{see the supplementary materials for the derivation \cite{SupplMat}}. The formula shows how the radial dependence of the cavity resonance is superimposed on the Gaussian envelope of the injected beam. As a result, changing the laser detuning \textcolor{black}{is to change the position of constructive interference in the intra-cavity field}.  \textcolor{black}{We also calculated the effect of the wavefronts and calculated the radial phase difference imprinted on the atoms, see Eq.~(\ref{eq:phase}) of \cite{SupplMat}, and demonstrated that it corresponds to the one of the Gaussian envelop of Eq.~(\ref{eq:static}), and is negligible.}

\begin{figure}
\centering
\includegraphics[width=1\linewidth]{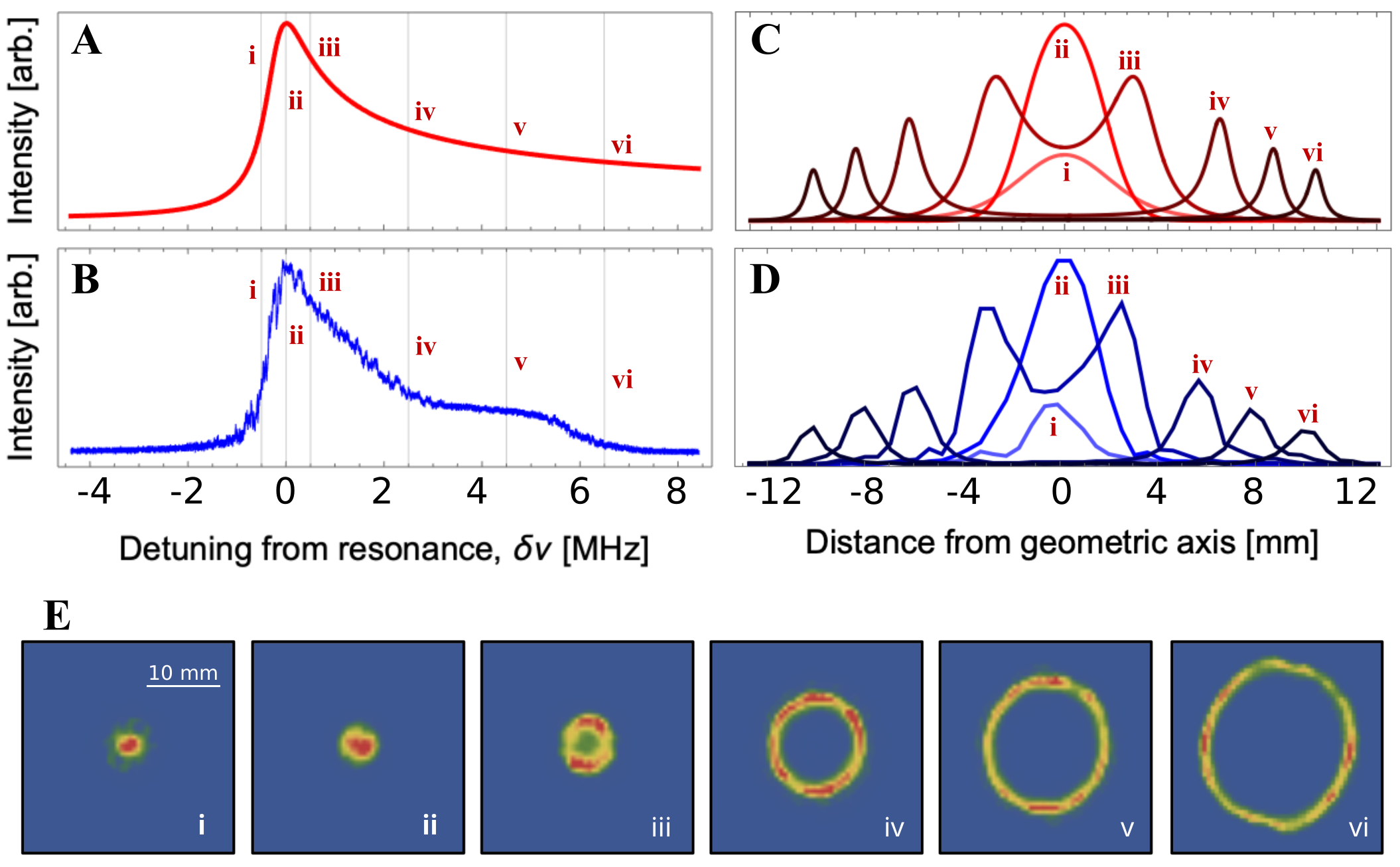}
\caption{\textbf{Frequency-to-space mapping in a marginally stable cavity.}
(A) Calculated resonance profile showing the excitation of transverse ring modes with increasing detuning \cite{Arnaud1970}. 
(B) Measured cavity transmission versus detuning, exhibiting the characteristic asymmetric line shape of an imperfect degenerate resonator. The signal drop at $\approx$ 6 MHz is due to beam clipping on the detection system, not in the cavity.
(C,D) Theoretical and experimental transverse cross sections, showing Gaussian-envelope ring structures. 
(E) CCD images of the intra-cavity field for fixed detunings, illustrating the continuous evolution from a central mode to expanding rings.
These measurements demonstrate that laser detuning directly maps onto the transverse spatial structure of the cavity field.
}
\label{fig:fig2}
\end{figure}

This frequency-to-space mapping is observed directly in the cavity transmission. The predicted and measured transmission spectrum at 780 nm is markedly asymmetric (Fig.~\ref{fig:fig2}-A,B), and cross-sections of the intra-cavity field show a continuous evolution from a central spot to rings of increasing radius as the detuning is increased (Fig.~\ref{fig:fig2}-C,D,E), in agreement with theoretical expectations for imperfect degenerate resonators \cite{Arnaud1970}. The measured spectra and transverse profiles agree well with the model and establish that the laser detuning provides direct control over the radius of the region of strongest intra-cavity coupling.

\section*{Scanning-cavity interrogation}

We exploit the radial resonance structure by applying a linear frequency chirp of rate $\delta\nu(t)=\alpha t$ across the cavity resonance. The resulting intra-cavity intensity experienced by the atoms at radius $r$ follows a Lorentzian temporal profile:
\begin{equation}
I(t,r)=
\frac{I_0(t)}{1+\left(\frac{t}{\tau}-\frac{r^2}{r_c^2}\right)^2}
\exp\!\left(-\frac{2r^2}{\omega_0^2}\right),
\label{eq:chirped}
\end{equation}
with $\tau=\Delta\nu_{\mathrm{FSR}}/(2F \alpha)$ and $r_c=\sqrt{\lambda f^2/(2\delta_1 F)}$.

At a given time, the intensity is maximal at the radius $r_b(t)=r_c\sqrt{t/\tau}$, so that the chirp dynamically sweeps the interaction region from the cavity axis to larger radii. A key feature of this response is that the temporal width of the local Lorentzian profile in Eq.~(\ref{eq:chirped}) is equal to $2\tau$ and is independent of radius. Atoms located at different radii therefore experience the same effective pulse shape, shifted only in time according to when the moving bright ring crosses their position. This turns the detuning chirp into a controlled scan of the atom-light interaction across the cloud.

To compensate for the Gaussian envelope of the injected mode, we simultaneously shape the pulse amplitude by acting on the radio-frequency amplitude driving the AOM. Fig.~\ref{fig:fig3}-A shows the power at the cavity output as a function of the frequency detuning, which is swept from -3 to +9 MHz. The power is switched off at a detuning of 8 MHz. The combination of chirp and amplitude shaping transforms the strongly inhomogeneous instantaneous cavity field into an effective extended interaction region with nearly uniform coupling over the scanned area.

Experimentally, we observe that this protocol produces a major improvement in Bragg diffraction efficiency. Compared with static excitation, the chirped pulse increases the maximum Rabi transfer efficiency by about a factor of 6 from 5\% to 28\% (Fig.~\ref{fig:fig3}-B). The optimum is obtained for a pulse duration of 130 $\mu$s corresponding to a chirp rate $\alpha =$ 92 kHz/$\mu s$. From the response of the full ensemble, we infer an effective interaction diameter of roughly 15 mm FWHM, with a transfer efficiency of about 28\%, substantially beyond the expected value of 14\% for a free-space Gaussian beam of the same diameter addressing the present cloud. This is made possible by compensating for the drop in power at the edge of the cavity.

\section*{Scanning-cavity interferometer}

\begin{figure}
\centering
\includegraphics[width=1\linewidth]{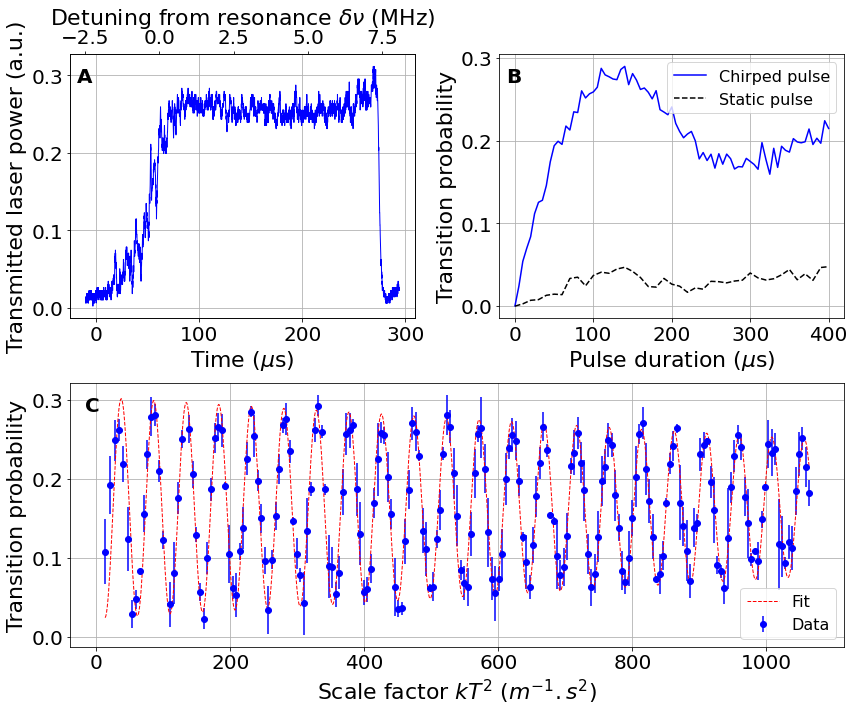}
\caption{\textbf{Scanning-cavity atom interferometry.}
(A) Power modulation at the cavity output when applying the frequency-chirped Bragg pulse (see Fig.~\ref{fig:fig2}-B) from 0 to 7.5 MHz relative to the cavity resonance at its center position. 
(B) Transition probability as a function of the total pulse duration for static excitation (gray dashed) and chirped excitation with the pulse described in Fig. A (blue). 
(C) Interferometer signal obtained using chirped pulses for a total interrogation time up to 16 ms. 
\label{fig:fig3}}
\end{figure}

Finally, we realize a three-pulse interferometer separated by an equal time interval $T$. When neglecting the pulse durations, the phase shift $\phi_{\rm{at}}^0$ of such interferometer is:
\begin{equation}
\phi_{\rm{at}}^{0} = k_{\rm{eff}} \cdot a \cdot T^2,
\label{eq:phase-all}
\end{equation}
where $k_{\rm{eff}}=4\pi \times \nu/c$ is the effective wave vector of the two-photon Bragg transition and $a$ the acceleration along the Bragg direction. The three pulses are chirped with the same chirp rate, and with twice more power in the case of the center $\pi$ pulse than for the first and last $\pi$/2 pulses. Fig.~\ref{fig:fig3}-C shows the resulting interferometer fringes as a function of the scale factor $k_{\rm{eff}} T^2$. The period of the fringes is related to the angle $\theta$ of the cavity compared to the horizontal direction, about 13 mrad, leading to an acceleration $a$ due to the projection of gravity $g$ along the direction of the cavity $a=g \cdot \sin(\theta)$. 
A contrast of 24\% is obtained for short interaction times. This shows that, in this chirped protocol, all atoms participating to the Rabi oscillation contribute as well to the interferometer signal in a coherent manner. When $T$ is very small, the displacement of atoms in the cavity during the interferometer can indeed be neglected. Therefore, the resonance frequency $\nu$, effective wave vector $k_{\rm{eff}}$ and time interval $T$ remain the same for each atom. 

As $T$ increases, one must take into account the displacements of atoms between pulses in the transverse directions of the cavity. These displacements are due to the acceleration by gravity and the  velocity dispersion of the atomic cloud. Consequently, in each pulse, numbered $i$ from 1 to 3, an atom experiences a different resonance condition $\nu_i=\nu+\delta \nu_i$, fulfilled for a different value of $k^i_{\rm{eff}}=k_{\rm{eff}}+\delta k_i$ (with $\delta k_i /k_{\rm{eff}}=\delta \nu_i /\nu$) and a different time of the pulse $t_i=(i-1)T+\delta t_i$ \textbf (with $\delta t_i=\delta \nu_i/\alpha$). In the limit of a high finesse cavity, the phase difference between the two counter propagating Bragg beams inside the cavity at the position of the atoms, at a distance $z$ compared to the mirror $M_1$, depends only on $k_{\rm{eff}}^i$ at the moment $t_i$ of the pulse. Here, aberrations of the mirror $M_1$ are neglected. Using the formalism derived in \cite{Antoine2003} and assuming that the phase imprinted on the diffracted wave packet corresponds to the optical phase difference between the Bragg beams at the moment of maximum intensity and for small frequency variations ($\delta\nu_i \ll \nu$), we can express the phase shift of the interferometer to first-order in $\delta\nu_i$ (or $\delta k_i$ and $\delta t_i$) as:

\begin{equation}
\begin{split} 
\phi_{\rm{at}} \sim & \, k_{\rm{eff}} \cdot a \cdot T^2 \left(1+\frac{\delta k_2}{k_{\rm{eff}}}+\frac{\delta t_3-\delta t_1}{T}\right) \\
&+z \cdot (\delta k_1-2\delta k_2+\delta k_3).
\end{split}
\label{eq:phase-dnu}
\end{equation}

The phase shift is calculated for a zero mean velocity at time $t_2$, corresponding to the center of the interferometer (\textit{i.e.} the $\pi$ pulse). We first identify a modification of the scale factor associated with the acceleration term $\phi_{\rm{at}}^0$ arising from variations in $k_{\rm{eff}}$ and from shifts in the pulse timings $\delta t_i$. In addition, we find a bias term associated with the atom-mirror distance $z$, which results from the variation of the transition wave vector $k_{\rm{eff}}$ between the three pulses.

For our experimental parameters, namely a distance $z = 20$ cm and an interrogation time $2T=16$ ms, the atomic displacement during the pulse is of the order of 0.3 mm. This allows us to estimate the corresponding impact on the contrast for the two contributions considered. For the bias term associated with the atom-mirror distance, the atomic velocity dispersion at $2T= 16$ ms induces a dephasing of approximately $2.5$ mrad, while the transverse displacement of the atoms along their ballistic trajectory induces a dephasing of approximately $4$ mrad. The scale factor inhomogeneity is estimated to be about $0.1 \% $, corresponding to a dephasing of approximately $60$ mrad at $2T= 16$ ms. These values are too small to account for the observed reduction in contrast with increasing interrogation time, as shown in Fig.~\ref{fig:fig3}-C. A possible explanation for the observed contrast reduction is the presence of vibrations, in particular relative displacements between the different optical elements in the cavity.

\section*{Conclusion}

In conclusion, we have shown that the spatial inhomogeneity of a marginally stable optical cavity can be exploited as a resource for cavity-enhanced atom interferometry. By chirping the interrogation frequency across the cavity ring resonances and synchronizing the chirp with amplitude modulation, we realize an effectively homogeneous single frequency Bragg diffraction over an extended area that is inaccessible under static excitation. This approach strongly enhances the diffraction efficiency while preserving the coherence required for atom interferometry. Our results establish dynamic excitation of marginally stable cavities as a practical route toward large-area, cavity-enhanced atom interferometers. 

Following this initial demonstration, a detailed investigation of vibration-induced dephasing across the cavity, as well as of the relative displacements of the lens and the input mirror $M_2$ with respect to the reference mirror $M_1$, should enable significantly longer interrogation times. In the near horizontal interrogation geometry, the dominant decoherence mechanism is expected to arise from atom-to-atom variations of the effective wave vector $k_{\rm{eff}}$. This contribution could be mitigated by bringing the atomic cloud closer to the mirror and, ultimately, compensated for by applying a small displacement to the cavity mirror $M_1$ with the PZT synchronously with the frequency scan.

Beyond atom interferometry, this technique also offers considerable potential for engineering tailored dipole traps in atom-optics experiments, including box-shaped and ring-shaped geometries. Its intrinsic stability and compatibility with dynamic modulation provide a versatile platform for shaping and controlling optical potentials.  
\medskip





The authors warmly thank J. Junca, I. Riou, G. Lef\`{e}vre, N. Mielec, and R. Geiger for their early work on the experiment, and F. Pereira Dos Santos for enlightening discussions.
We acknowledge the engineering support of P. Teulat, L. Sidorenkov, J. Pinto, and M. Lours.
\paragraph*{Funding:}

This work was supported by the French State through the “Agence Nationale de la Recherche” (ANR) within the framework of the “Investissement d'Avenir” programs Equipex MIGA (ANR-11-EQPX-0028), IdEx Bordeaux - LAPHIA (ANR-10-IDEX-03-02) and the projects PIMAI ANR-18-CE47-0002-01 and EOSBECMR ANR-18-CE91-0003-01.
We acknowledge financial support from the European Union’s Horizon 2020 research and innovation program (FET-Open No. 964531 CRYST$^3$, H2020 Qu-Test N. 101113901), Euramet (Project 23FUN02 CoCoRICO), the EUR Light S\&T Graduate Program (PIA3 Program “Investment for the Future”, ANR-17-EURE-0027), the Idex of the University of Bordeaux (Research Program "GPR Light"), Ville de Paris (project HSENS-MWGRAV). Qu-Test has received funding from the European Union’s Horizon Europe research and innovation program under the Grant Agreement 101113901. We acknowledge support from the NAQUIDIS quantum center.

\paragraph*{Author contributions:}
DOS carried out the experiment and analysed the data with QB and BC.
AB, QB, and AL developed the formalism to interpret the data, with input from DOS and BC.
All authors participated in writing the manuscript.
AL, PB, and BC supervised the research.
The authors attest that this manuscript was written without the aid or consultation of any form of artificial intelligence (AI) or large language model (LLM).

\paragraph*{Competing interests:}
The authors declare that they have no competing interests.
\paragraph*{Data and materials availability:}
All data needed to evaluate the conclusions of this work are present in the text, supplementary material, or are available from the authors upon reasonable request.

\bibliography{biblio}

@article{Hamilton2015,
  doi = {10.1103/physrevlett.114.100405},
  url = {https://doi.org/10.1103/physrevlett.114.100405},
  year = {2015},
  month = mar,
  publisher = {American Physical Society ({APS})},
  volume = {114},
  number = {10},
  pages = {100405},
  author = {Paul Hamilton and Matt Jaffe and Justin M. Brown and Lothar Maisenbacher and Brian Estey and Holger M\"{u}ller},
  title = {Atom Interferometry in an Optical Cavity},
  journal = {Phys. Rev. Lett.}
}

@article{Mielec2020,
author = {Nicolas Mielec and Ranjita Sapam and Constance Poulain and Arnaud Landragin and Andrea Bertoldi and Philippe Bouyer and Benjamin Canuel and Remi Geiger},
journal = {Opt. Expr.},
number = {26},
pages = {39112--39127},
publisher = {OSA},
title = {Degenerate optical resonator for the enhancement of large laser beams},
volume = {28},
month = {Dec},
year = {2020},
url = {http://www.opticsexpress.org/abstract.cfm?URI=oe-28-26-39112},
doi = {10.1364/OE.409293}
}

@article{Sabulsky2024,
  title = {Multiphoton Atom Interferometry via Cavity-Enhanced {B}ragg Diffraction},
  author = {Sabulsky, D. O. and Junca, J. and Zou, X. and Bertoldi, A. and Prevedelli, M. and Beaufils, Q. and Geiger, R. and Landragin, A. and Bouyer, P. and Canuel, B.},
  collaboration = {MIGA Consortium},
  journal = {Phys. Rev. Lett.},
  volume = {132},
  issue = {21},
  pages = {213601},
  numpages = {7},
  year = {2024},
  month = {May},
  publisher = {American Physical Society},
  doi = {10.1103/PhysRevLett.132.213601},
  url = {https://link.aps.org/doi/10.1103/PhysRevLett.132.213601}
}

@article{Arnaud1969,
  title = {Degenerate Optical Cavities},
  volume = {8},
  ISSN = {1539-4522},
  url = {http://dx.doi.org/10.1364/AO.8.000189},
  DOI = {10.1364/ao.8.000189},
  number = {1},
  journal = {Appl. Opt.},
  publisher = {Optica Publishing Group},
  author = {Arnaud,  J. A.},
  year = {1969},
  month = jan,
  pages = {189}
}

@article{Riou2017,
  author={I Riou and N Mielec and G Lefèvre and M Prevedelli and A Landragin and P Bouyer and A Bertoldi and R Geiger and B Canuel},
  title={A marginally stable optical resonator for enhanced atom interferometry},
  journal={J. Phys. B},
  volume={50},
  number={15},
  pages={155002},
  url={http://stacks.iop.org/0953-4075/50/i=15/a=155002},
  year={2017},
}

@phdthesis{Junca2022.PhD,
  TITLE = {Progress of the {MIGA} project toward gravity strain measurements with atom interferometry},
  AUTHOR = {Junca, Joseph},
  URL = {https://tel.archives-ouvertes.fr/tel-03669058},
  NUMBER = {2022BORD0150},
  SCHOOL = {Bordeaux University},
  YEAR = {2022},
  MONTH = Apr,
  TYPE = {{Ph.D. thesis}},
  HAL_ID = {tel-03669058},
  HAL_VERSION = {v1},
}

@article{Arnaud1970,
author = {J. A. Arnaud},
journal = {Appl. Opt.},
number = {5},
pages = {1192--1200},
publisher = {Optica Publishing Group},
title = {Degenerate {O}ptical {C}avities. {III}: {E}ffect of {A}berrations},
volume = {9},
month = {May},
year = {1970},
url = {https://opg.optica.org/ao/abstract.cfm?URI=ao-9-5-1192},
doi = {10.1364/AO.9.001192}
}

@article{Chiow2011,
  title = {102$\hbar$k Large Area Atom Interferometers},
  volume = {107},
  ISSN = {1079-7114},
  url = {http://dx.doi.org/10.1103/PhysRevLett.107.130403},
  DOI = {10.1103/physrevlett.107.130403},
  number = {13},
  pages = {130403},
  journal = {Phys. Rev. Lett.},
  publisher = {American Physical Society (APS)},
  author = {Chiow,  Sheng-wey and Kovachy,  Tim and Chien,  Hui-Chun and Kasevich,  Mark A.},
  year = {2011},
  month = sep 
}

@article{Rodzinka2024,
  title = {Optimal {F}loquet state engineering for large scale atom interferometers},
  volume = {15},
  ISSN = {2041-1723},
  url = {http://dx.doi.org/10.1038/s41467-024-54539-w},
  DOI = {10.1038/s41467-024-54539-w},
  number = {1},
  pages = {10281},
  journal = {Nat. Commun.},
  publisher = {Springer Science and Business Media LLC},
  author = {Rodzinka,  T. and Dionis,  E. and Calmels,  L. and Beldjoudi,  S. and Béguin,  A. and Guéry-Odelin,  D. and Allard,  B. and Sugny,  D. and Gauguet,  A.},
  year = {2024},
  month = nov 
}

@article{Dickerson2013,
  title = {Multiaxis Inertial Sensing with Long-Time Point Source Atom Interferometry},
  volume = {111},
  ISSN = {1079-7114},
  url = {http://dx.doi.org/10.1103/PhysRevLett.111.083001},
  DOI = {10.1103/physrevlett.111.083001},
  number = {8},
  pages = {083001},
  journal = {Phys. Rev. Lett.},
  publisher = {American Physical Society (APS)},
  author = {Dickerson,  Susannah M. and Hogan,  Jason M. and Sugarbaker,  Alex and Johnson,  David M. S. and Kasevich,  Mark A.},
  year = {2013},
  month = aug 
}

@misc{SupplMat,
  note         = "See Supplemental Material for derivation of the resonant forward and backward propagating EM fields inside the cavity and the expression of their transverse phase difference."
}

@article{Kogelnik1966,
  title = {Laser Beams and Resonators},
  volume = {5},
  ISSN = {1539-4522},
  url = {http://dx.doi.org/10.1364/AO.5.001550},
  DOI = {10.1364/ao.5.001550},
  number = {10},
  journal = {Appl. Opt.},
  publisher = {Optica Publishing Group},
  author = {Kogelnik,  H. and Li,  T.},
  year = {1966},
  month = oct,
  pages = {1550}
}

@article{Panda2024,
  title = {Coherence limits in lattice atom interferometry at the one-minute scale},
  volume = {20},
  ISSN = {1745-2481},
  url = {http://dx.doi.org/10.1038/s41567-024-02518-9},
  DOI = {10.1038/s41567-024-02518-9},
  number = {8},
  journal = {Nat. Phys.},
  publisher = {Springer Science and Business Media LLC},
  author = {Panda,  Cristian D. and Tao,  Matthew and Egelhoff,  James and Ceja,  Miguel and Xu,  Victoria and M\"{u}ller,  Holger},
  year = {2024},
  month = june,
  pages = {1234–1239}
}

@article{Gautier2022,
  title = {Accurate measurement of the Sagnac effect for matter waves},
  volume = {8},
  ISSN = {2375-2548},
  url = {http://dx.doi.org/10.1126/sciadv.abn8009},
  DOI = {10.1126/sciadv.abn8009},
  number = {23},
  pages = {eabn8009},
  journal = {Sci. Adv.},
  publisher = {American Association for the Advancement of Science (AAAS)},
  author = {Gautier,  Romain and Guessoum,  Mohamed and Sidorenkov,  Leonid A. and Bouton,  Quentin and Landragin,  Arnaud and Geiger,  Remi},
  year = {2022},
  month = June 
}

@article{Long2007,
author = {R. Long and A. K. Tuchman and M. A. Kasevich},
journal = {Opt. Lett.},
number = {17},
pages = {2502--2504},
publisher = {Optica Publishing Group},
title = {Multiple frequency modulation for low-light atom measurements in an optical cavity},
volume = {32},
month = {Sep},
year = {2007},
url = {https://opg.optica.org/ol/abstract.cfm?URI=ol-32-17-2502},
doi = {10.1364/OL.32.002502},
}

@article{schuldt2015,
  title = {Design of a dual species atom interferometer for space},
  author = {Schuldt, Thilo and Schubert, Christian and Krutzik, Markus and Bote, Lluis Gesa and Gaaloul, Naceur and Hartwig, Jonas and Ahlers, Holger and Herr, Waldemar and {Posso-Trujillo}, Katerine and Rudolph, Jan and Seidel, Stephan and Wendrich, Thijs and Ertmer, Wolfgang and Herrmann, Sven and {Kubelka-Lange}, Andr{\'e} and Milke, Alexander and Rievers, Benny and Rocco, Emanuele and Hinton, Andrew and Bongs, Kai and Oswald, Markus and Franz, Matthias and Hauth, Matthias and Peters, Achim and Bawamia, Ahmad and Wicht, Andreas and Battelier, Baptiste and Bertoldi, Andrea and Bouyer, Philippe and Landragin, Arnaud and Massonnet, Didier and L{\'e}v{\`e}que, Thomas and Wenzlawski, Andre and Hellmig, Ortwin and Windpassinger, Patrick and Sengstock, Klaus and Von Klitzing, Wolf and Chaloner, Chris and Summers, David and Ireland, Philip and Mateos, Ignacio and Sopuerta, Carlos F. and Sorrentino, Fiodor and Tino, Guglielmo M. and Williams, Michael and Trenkel, Christian and Gerardi, Domenico and Chwalla, Michael and Burkhardt, Johannes and Johann, Ulrich and Heske, Astrid and Wille, Eric and Gehler, Martin and Cacciapuoti, Luigi and G{\"u}rlebeck, Norman and Braxmaier, Claus and Rasel, Ernst},
  year = 2015,
  month = jun,
  journal = {Experimental Astronomy},
  volume = {39},
  number = {2},
  pages = {167--206},
  issn = {0922-6435, 1572-9508},
  doi = {10.1007/s10686-014-9433-y},
  urldate = {2025-04-08},
 }

@article{gebbe2021,
  title = {Twin-lattice atom interferometry},
  author = {Gebbe, Martina and Siem{\ss}, Jan-Niclas and Gersemann, Matthias and M{\"u}ntinga, Hauke and Herrmann, Sven and L{\"a}mmerzahl, Claus and Ahlers, Holger and Gaaloul, Naceur and Schubert, Christian and Hammerer, Klemens and Abend, Sven and Rasel, Ernst M.},
  year = 2021,
  month = may,
  journal = {Nature Communications},
  volume = {12},
  number = {1},
  pages = {2544},
  issn = {2041-1723},
  doi = {10.1038/s41467-021-22823-8},
  urldate = {2025-02-10},
  }

@article{wilkason2022,
  title = {Atom {{Interferometry}} with {{Floquet Atom Optics}}},
  author = {Wilkason, Thomas and Nantel, Megan and Rudolph, Jan and Jiang, Yijun and Garber, Benjamin E. and Swan, Hunter and Carman, Samuel P. and Abe, Mahiro and Hogan, Jason M.},
  year = 2022,
  month = oct,
  journal = {Physical Review Letters},
  volume = {129},
  number = {18},
  pages = {183202},
  issn = {0031-9007, 1079-7114},
  doi = {10.1103/PhysRevLett.129.183202},
  urldate = {2025-02-07},
  
}

@article{Greve2022,
  title = {Entanglement-enhanced matter-wave interferometry in a high-finesse cavity},
  volume = {610},
  ISSN = {1476-4687},
  url = {http://dx.doi.org/10.1038/s41586-022-05197-9},
  DOI = {10.1038/s41586-022-05197-9},
  number = {7932},
  journal = {Nature},
  publisher = {Springer Science and Business Media LLC},
  author = {Greve,  Graham P. and Luo,  Chengyi and Wu,  Baochen and Thompson,  James K.},
  year = {2022},
  month = oct,
  pages = {472–477}
}

@article{Zhan2019,
  doi = {10.1142/s0218271819400054},
  url = {https://doi.org/10.1142/s0218271819400054},
  year = {2019},
  month = may,
  publisher = {World Scientific Pub Co Pte Lt},
  pages = {1940005},
  author = {Ming-Sheng Zhan and Jin Wang and Wei-Tou Ni and Dong-Feng Gao and Gang Wang and Ling-Xiang He and Run-Bing Li and Lin Zhou and Xi Chen and Jia-Qi Zhong and Biao Tang and Zhan-Wei Yao and Lei Zhu and Zong-Yuan Xiong and Si-Bin Lu and Geng-Hua Yu and Qun-Feng Cheng and Min Liu and Yu-Rong Liang and Peng Xu and Xiao-Dong He and Min Ke and Zheng Tan and Jun Luo},
  title = {{ZAIGA}: Zhaoshan long-baseline atom interferometer gravitation antenna},
  journal = {International Journal of Modern Physics D}
}

@article{Canuel2018,
  doi = {10.1038/s41598-018-32165-z},
  url = {https://doi.org/10.1038/s41598-018-32165-z},
  year  = {2018},
  month = {sep},
  publisher = {Springer Nature America,  Inc},
  volume = {8},
  number = {1},
  pages = {14064},
  author = {B. Canuel and A. Bertoldi and L. Amand and E. Pozzo di Borgo and T. Chantrait and C. Danquigny and M. Dovale {\'{A}}lvarez and B. Fang and A. Freise and R. Geiger and J. Gillot and S. Henry and J. Hinderer and D. Holleville and J. Junca and G. Lef{\`{e}}vre and M. Merzougui and N. Mielec and T. Monfret and S. Pelisson and M. Prevedelli and S. Reynaud and I. Riou and Y. Rogister and S. Rosat and E. Cormier and A. Landragin and W. Chaibi and S. Gaffet and P. Bouyer},
  title = {Exploring gravity with the {MIGA} large scale atom interferometer},
  journal = {Sci. Rep.}
}

@article{Canuel2020,
	doi = {10.1088/1361-6382/aba80e},
	url = {https://doi.org/10.1088/1361-6382/aba80e},
	year = 2020,
	month = {oct},
	publisher = {{IOP} Publishing},
	volume = {37},
	number = {22},
	pages = {225017},
	author = {B Canuel and S Abend and P Amaro-Seoane and F Badaracco and Q Beaufils and A Bertoldi and K Bongs and P Bouyer and C Braxmaier and W Chaibi and N Christensen and F Fitzek and G Flouris and N Gaaloul and S Gaffet and C L Garrido Alzar and R Geiger and S Guellati-Khelifa and K Hammerer and J Harms and J Hinderer and M Holynski and J Junca and S Katsanevas and C Klempt and C Kozanitis and M Krutzik and A Landragin and I L{\`{a}}zaro Roche and B Leykauf and Y-H Lien and S Loriani and S Merlet and M Merzougui and M Nofrarias and P Papadakos and F Pereira dos Santos and A Peters and D Plexousakis and M Prevedelli and E M Rasel and Y Rogister and S Rosat and A Roura and D O Sabulsky and V Schkolnik and D Schlippert and C Schubert and L Sidorenkov and J-N Siem{\ss} and C F Sopuerta and F Sorrentino and C Struckmann and G M Tino and G Tsagkatakis and A Vicer{\'{e}} and W von Klitzing and L Woerner and X Zou},
	title = {{ELGAR}{\textemdash}a {E}uropean {L}aboratory for {G}ravitation and {A}tom-interferometric {R}esearch},
	journal = {Classical and Quantum Gravity},
	
}

@article{Beaufils2022,
  doi = {10.1038/s41598-022-23468-3},
  url = {https://doi.org/10.1038/s41598-022-23468-3},
  year = {2022},
  month = nov,
  publisher = {Springer Science and Business Media {LLC}},
  volume = {12},
  number = {1},
  author = {Quentin Beaufils and Leonid A. Sidorenkov and Pierre Lebegue and Bertrand Venon and David Holleville and Laurent Volodimer and Michel Lours and Joseph Junca and Xinhao Zou and Andrea Bertoldi and Marco Prevedelli and Dylan O. Sabulsky and Philippe Bouyer and Arnaud Landragin and Benjamin Canuel and Remi Geiger},
  title = {Cold-atom sources for the Matter-wave laser Interferometric Gravitation Antenna ({MIGA})},
  journal = {Scientific Reports}
}

@article{Antoine2003,
  title = {Exact phase shifts for atom interferometry},
  author = {Antoine, {\relax Ch}. and Bord{\'e}, {\relax Ch.J}.},
  year = 2003,
  month = jan,
  journal = {Physics Letters A},
  volume = {306},
  number = {5-6},
  pages = {277--284},
  issn = {03759601},
  doi = {10.1016/S0375-9601(02)01625-0},
  urldate = {2026-05-25},
}

@article{Erden1997,
  doi = {10.1364/josaa.14.002190},
  url = {http://dx.doi.org/10.1364/JOSAA.14.002190},
  year  = {1997},
  month = {sep},
  publisher = {The Optical Society},
  volume = {14},
  number = {9},
  pages = {2190-2194},
  author = {M. Fatih Erden and Haldun M. Ozaktas},
  title = {Accumulated {G}ouy phase shift in {G}aussian beam propagation through first-order optical systems},
  journal = {JOSA A}
}

@Book{Siegman1986,
  Title                    = {Lasers},
  Author                   = {Siegman, A.~E.},
  Publisher                = {University Science Books},
  Year                     = {1986},
  ISBN                     = {9780935702118},
  Lccn                     = {81051269},
  Url                      = {https://books.google.fr/books?id=1BZVwUZLTkAC}
}

\newpage

\onecolumngrid

\newpage

\section*{Supplementary Material}

\begin{figure}[h]
\centering
\includegraphics[width=0.5\linewidth]{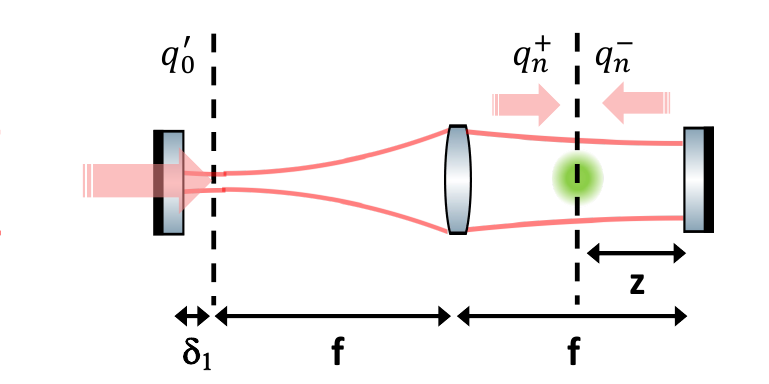}
\caption{Resonator geometry considered for the intra-cavity field calculation. $q_n^{\pm}$ are the complex radius of curvature of the Gaussian beam at position z in the forward (+) and backwards (-) directions after $n$ round trips inside the cavity. $q'_0$ is the complex radius if curvature of the input Gaussian field.
}
\label{fig:cavityconfig}
\end{figure}

\noindent 
In the following, we calculate the resonant forward and backward propagating EM fields inside the cavity and derive the expression of their transverse phase difference. We will consider the situation described in Fig. \ref{fig:cavityconfig}: an input Gaussian field on the resonator $q'_0 = iz_{r_0}^{'}$ located at the focal point of the intra-cavity lens, and a confocality error of $\delta_1$ in the longitudinal alignment of the input mirror. 

Note that $q_n^{\pm}$ is the complex radius of curvature of the Gaussian beam at position z in the forward (+) and backwards (-) directions after $n$ round trips inside the cavity. The beam size $\omega_{n}^{\pm}$ and the real radius of curvature $R_{n}^{\pm}$ of the Gaussian beam at position z in the forward (+) and backwards (-) directions after n round trips can thus be derived from:
\begin{equation}
\left \{
\begin{array}{c c l}
\omega_{n}^{\pm}&=&\sqrt[]{\frac{-\lambda}{\pi\,\,\mathrm{Im}(1/q_n^{\pm})}}  \\
1/R_{n}^{\pm}&=&\mathrm{Re}(1/q_n^{\pm}).
\end{array}
\right.
\end{equation}
\subsection{Injected beam parameters at position z}
The complex radius of curvature of the injected beam at position z, $q_0^{+}$, can be obtained from the input beam $q'_0 = iz_{r_0}^{'}$ by using the ABCD transfer matrix \cite{Kogelnik1966}:

\begin{equation}
U=
\begin{pmatrix}
1&+f-z\\
0&1
\end{pmatrix}
\begin{pmatrix}
1&0\\
-1/f&1
\end{pmatrix}
\begin{pmatrix}
1&f\\
0&1
\end{pmatrix}.
\end{equation}
We therefore have:
\begin{equation}
1/q_0^{+}  = \frac{U_{22}/q'_0+U_{21}}{U_{12}/q'_0+U_{11}},
\end{equation}
leading to:
\begin{equation}
1/q_0^{+}  =  -i\frac{z_{r_0}^{'}}{f^2}- \frac{zz_{r_0}^{'2}}{f^4}.
\end{equation}
Taking the Gouy phase reference at location of the incoming beam, we obtain the following incoming beam parameters at position z:

\begin{equation}
\left \{
\begin{array}{c c c}
\omega_{0}^{+}&=&\frac{\lambda f} {\pi \omega_0^{'}} \\
1/R_{0}^{+}&=&- \frac{zz_{r_0}^{'2}}{f^4} \\
\Phi_{G0}^{+}&=&0.
\end{array}
\right.
\end{equation}

\subsection{Calculation of the forward propagating resonant EM field}
\noindent The transfer matrix for one round trip inside the resonator from position z can be expressed as:

\begin{equation}
V=
\begin{pmatrix}
1&f-z\\
0&1
\end{pmatrix}
\begin{pmatrix}
1&0\\
-1/f&1
\end{pmatrix}
\begin{pmatrix}
1&2(\delta_{1}+f)\\
0&1
\end{pmatrix}
\begin{pmatrix}
1&0\\
-1/f&1
\end{pmatrix}
\begin{pmatrix}
1&z+f\\
0&1
\end{pmatrix}=\left(
\begin{array}{cc}
 1 -\epsilon z& -\epsilon z^2 \\
\epsilon & -1 +\epsilon z\\
\end{array}
\right),
\end{equation}
with:
\begin{equation}
\epsilon =  \frac{2 \delta_1 z}{f^2}.
\end{equation}
This enables to calculate the transfer matrix for n round trips can from the matrix power:

\begin{equation}
V^n=(-1)^n\left(
\begin{array}{cc}
 1 +n\epsilon z& n\epsilon z^2 \\
-n\epsilon & 1 -n\epsilon z\\
\end{array}
\right)
\end{equation}
The complex radius of curvature after n round trips $1/q_n^{+}$ of the forward beam is obtained from the injected beam $1/q_0^{+}$ using the $V^n$ matrix coefficients:

\begin{equation}
1/q_n^{+}  = \frac{V_{22}^n/q_0^{+}+V_{21}^n}{V_{12}^n/q_0^{+}+V_{11}^n}. 
\end{equation}
Neglecting third order terms in $\epsilon z_{r_0}^{'2}$, we have: 

\begin{equation}
1/q_n^{+}  = -i(1-2n\epsilon z)\frac{z_{r_0}^{'}}{f^2}-\frac{zz_{r_0}^{'2}}{f^4}-n\epsilon,
\end{equation}
and we obtain the following beam parameters at position z after n round trips:
\begin{equation}\label{param+}
\left \{
\begin{array}{c c c}
\omega_{n}^{+}&=&\omega_{0}^{+} (1+n\epsilon) \\
1/R_{n}^{+}&=&-n\epsilon- \frac{zz_{r_0}^{'2}}{f^4}, \\
\end{array}
\right.
\end{equation}
and an accumulated Gouy phase \cite{Erden1997}:

\begin{equation}\label{phig+}
\phi_{Gn}^{+} = \phi_{G0}^{+} + \tan^{-1}\left( \frac{\lambda V_{12}^n}{\left(V_{11}^n+\frac{V_{12}^n}{R_{0}^+}\right)\pi \omega_{0}^{+2}} \right) =n\epsilon z^2\frac{z_{r_0}^{'}}{f^2}.
\end{equation}
This leads to the following expression \cite{Siegman1986} for the forward EM field after n round trips:
\begin{equation}
E_n^+(r,z) = E_0^+\rho^{n}
e^{ikz}e^{-2ikn L}e^{i\phi_{Gn}^+} \frac{1}{\omega_n^+}e^{-\frac{r^2}{\omega_n^{+2}}}e^{-i\frac{kr^2}{2R_n^+}},
\end{equation}
and so the total forward EM field can then be expressed: 
\begin{equation}
E^+(r,z) =\sum_{n=0}^{\infty}E_n^+(r,z).
\end{equation}
Using Eq. \ref{param+} and \ref{phig+}, we obtain: 
\begin{equation}\label{sum+}
E^+(r,z)  = E_0^+(r)e^{ikz} e^{i\frac{kr^2}{2}\frac{zz_{r_0}^{'2}}{f^4}}\sum_{n=0}^{\infty}(1-n\epsilon z\alpha) (\rho e^{i\varphi})^{n},
\end{equation}
with:
\begin{equation}
\left \{
\begin{array}{c c c}
\alpha&=&(1-\frac{r^2}{\omega_{0}^{+2}}) \\
\varphi&=&-2kL+\epsilon\frac{z^2}{z_{r_0}^{+}}+\epsilon\frac{kr^2}{2}\approx2kL+\epsilon\frac{kr^2}{2} \\
E_0^+(r)&=&\frac{E_0^+}{\omega_0^+}e^{-\frac{r^2}{\omega_0^{+2}}}.
\end{array}
\right.
\end{equation}
Using the power series identities $\sum_{n=0}^{\infty}u^n=\frac{1}{1-u}$ and $\sum_{n=0}^{\infty}nu^n=\frac{z}{(1-u)^2}$, Eq. \ref{sum+} becomes: 
\begin{equation}
E^+(r,z)  = E_0^+(r)e^{ikz} e^{i\frac{kr^2}{2}\frac{zz_{r_0}^{'2}}{f^4}} \frac{1-(1+\epsilon z\alpha)\rho e^{i\varphi}}{(1-\rho e^{i\varphi})^2}
\end{equation}
For the sake of simplicity, we will now discuss the resonance condition and intensity distribution of the cavity mode at position $z=0$ on mirror $M_2$. We obtain:
\begin{equation}
I^+(r)  = \frac{E_0^+E_0^{+*}}{\omega_0^{+2}}\frac{1}{1+\rho^2-2\rho\cos(\varphi) }e^{-\frac{2r^2}{\omega_0^{+2}}}.
\end{equation}
The resonance condition is thus given by $\varphi=2kL+\epsilon\frac{kr^2}{2}=2p\pi$. For a frequency $\delta\nu$ close to resonance, we obtain the following expression \cite{Junca2022.PhD}
\begin{equation}
I^+(\delta\nu,r) = \frac{I^+(0,0)}{1+4F^2\left(\frac{\delta\nu}{\Delta\nu_{\mathrm{FSR}}} - \frac{\delta_1 r^2}{\lambda f^2} \right)^2}e^{-\frac{2r^2}{\omega_0^{+2}}}.
\label{eq:app_static_intensity}
\end{equation}
This implies that for a given detuning $\delta\nu,$ the intra-cavity intensity is maximum at a resonant radius $r_b=\sqrt{\frac{\lambda f^2\delta\nu}{\delta_1\Delta\nu_{\mathrm{FSR}}}}$.
Assuming that this resonance condition remains unchanged at position z (an assumption we found to be valid in the limit of $\epsilon z F \ll 1$, which is verified considering our experimental parameters), the total forward field at position z and at the resonant radius $r_b$ can be written:
\begin{equation}\label{E+}
E^+(r_b,z)  = E_0^+(r_b)e^{ikz} e^{i\frac{kr_b^2}{2}\frac{zz_{r_0}^{'2}}{f^4}} \frac{1-(1+\epsilon z\alpha)\rho }{(1-\rho)^2}.
\end{equation}

\subsection{Calculation of the backward propagating resonant EM field}
\noindent The propagation Matrix after n rounds trips from the input beam $q_0^{+}$ to the backward propagating beam $q_n^{-}$ is:

\begin{equation}
W^n=\begin{pmatrix}
1&2z\\
0&1
\end{pmatrix}V^n=(-1)^n\left(
\begin{array}{cc}
 1 -n\epsilon z& 2z-n\epsilon z^2 \\
-n\epsilon & 1 -n\epsilon z\\
\end{array}
\right).
\end{equation}
We therefore obtain:
\begin{equation}
1/q_n^{-}  = \frac{W_{22}^n/q_0^{+}+W_{21}^n}{W_{12}^n/q_0^{+} W_{11}^n} = -i(1+2n\epsilon z)\frac{z_{r_0}^{'}}{f^2}+\frac{zz_{r_0}^{'2}}{f^4}-n\epsilon,
\end{equation}
leading to the following beam parameter and accumulated Gouy phase for the backward propagating beam:

\begin{equation}\label{param-}
\left \{
\begin{array}{c c l}
\omega_{n}^{-}&=&\omega_{0}^{+} (1-n\epsilon) \\
1/R_{n}^{-}&=&-n\epsilon+ \frac{zz_{r_0}^{'2}}{f^4}\\
\phi_{Gn}^{-} &= &\phi_{G0}^{+} + \tan^{-1}\left( \frac{\lambda W_{12}^n}{\left(W_{11}^n+\frac{W_{12}^n}{R_{0}^+}\right)\pi \omega_{0}^{+2}} \right) =n\epsilon z^2\frac{z_{r_0}^{'}}{f^2}+2z\frac{z_{r_0}^{'}}{f^2}.\\

\end{array}
\right.
\end{equation}
The backwards propagating EM field after n round trips can therefore be written:
\begin{equation}
E_n^-(r,z) = E_0^+\rho^{n}
e^{-ikz}e^{-2ikn L}e^{i\phi_{Gn}^-} \frac{1}{\omega_n^-}e^{-\frac{r^2}{\omega_n^{-2}}}e^{-i\frac{kr^2}{2R_n^-}},
\end{equation}
and so the total backward propagating EM field can then be expressed: 
\begin{equation}
E^-(r,z) =\sum_{n=0}^{\infty}E_n^-(r,z).
\end{equation}
Using the relations of  Eq. \ref{param-} we obtain: 
\begin{equation}
E^-(r,z)  = E_0^+(r)e^{-ikz} e^{-i\frac{kr^2}{2}\frac{zz_{r_0}^{'2}}{f^4}}e^{2iz\frac{z_{r_0}^{'}}{f^2}}\sum_{n=0}^{\infty}(1+n\epsilon z\alpha) (\rho e^{i\varphi})^{n}.
\end{equation}
Using the power series identities previously mentioned, we have:
\begin{equation}
E^-(r,z)  = E_0^+(r)e^{-ikz} e^{-i\frac{kr^2}{2}\frac{zz_{r_0}^{'2}}{f^4}}e^{2iz\frac{z_{r_0}^{'}}{f^2}} \frac{1-(1-\epsilon z\alpha)\rho e^{i\varphi}}{(1-\rho e^{i\varphi})^2},
\end{equation}
and the backward propagating field at position z at the resonant radius $r_b$ can thus be written:
\begin{equation}\label{E-}
E^-(r_b,z)  = E_0^+(r_b)e^{-ikz} e^{-i\frac{kr_b^2}{2}\frac{zz_{r_0}^{'2}}{f^4}}e^{2iz\frac{z_{r_0}^{'}}{f^2}}\frac{1-(1-\epsilon z\alpha)\rho }{(1-\rho)^2}.
\end{equation}

\subsection{Phase difference between total forward and backward propagating fields}
At position z and at the resonant radius $r_b$, the phase difference $\Delta\phi$ between total forward and backward propagating fields can be derived from Eq. \ref{E+} and \ref{E-}:
\begin{equation}
\Delta\phi(r_b,z)  = \arg(E^+(r_b,z))-\arg(E^-(r_b,z))=2kz+kr_b^2\frac{zz_{r_0}^{'2}}{f^4}-2z\frac{z_{r_0}^{'}}{f^2}=\Delta\phi(0,z)+kr_b^2\frac{zz_{r_0}^{'2}}{f^4}.
\end{equation}
\begin{equation}
\Delta\phi(r_b,z)=\Delta\phi(0,z)+kr_b^2\frac{1}{R_c^{0+}}.
\label{eq:phase}
\end{equation}
where $\frac{1}{R_c^{0+}}=\frac{z}{z_{r_0}^{+2}}$. The phase difference at resonant radius $r_b$ is thus twice the radial phase of a Gaussian beam of Rayleigh range $z_{r_0}^{+}$ at distance z from the waist, which is strictly negligible considering our experimental parameters ($z=20$ cm at the atom location and $z_{r_0}^{+}=$ 402 m).
\end{document}